# Role of energetic ions in the growth of fcc and ω crystalline phases in Ti films deposited by HiPIMS

D. Dellasega[1,*], F. Mirani[1], D. Vavassori[1], C. Conti[2] and M. Passoni[1]

[1]Politecnico di Milano, Via Ponzio 34/3, I-20133 Milano, Italy
[2] ISPC-CNR, Via Cozzi 54, 20125 Milano, Italy

*Corresponding author. E-mail: david.dellasega@polimi.it

## Abstract

In this work, we show how to control the nucleation of fcc and hexagonal (ω) crystalline phases in Ti films by adding a proper Ti ion flux to the film-forming species. To this aim, films with different thicknesses are grown by High Power Impulse Magnetron Sputtering (HiPIMS) and conventional Direct Current Magnetron Sputtering (DCMS) as a reference. HiPIMS depositions with different substrate bias voltage $U_S$ (0 V, -300 V and -500 V) are performed to investigate different ion energy ranges. Microstructure, morphology and residual stress of the deposited films, as well as the DCMS and HiPIMS plasma composition, are analysed with different characterization techniques. The DCMS samples, where the species are mostly atoms, exhibit the Ti α-phase only. As far as HiPIMS samples are concerned, films deposited in low energy ion conditions ($U_S$= 0 V) show the presence of the Ti fcc phase up to a maximum thickness of about 370 nm. Differently, films deposited under high energy conditions ($U_S$= -300 V and -500 V) show the nucleation of the Ti ω-phase for thicknesses greater than 260 and 330 nm, respectively. The formation of these unusual Ti phases is discussed considering the different deposition conditions.

## 1. Introduction

Titanium (Ti), thanks to its high mechanical strength, excellent thermal stability, good corrosion resistance in harsh conditions and intrinsic biocompatibility, has been exploited as bulk material in aerospace, industrial engineering and medical instrumentation fields [1]. The exceptional properties of Ti are also related to the different crystal structures that Ti, as well as transition metals of the IV group, exhibits depending on the process conditions [2]. Ti shows the hcp phase (named α-phase in the following) at room temperature under atmospheric pressure. On raising the temperature while keeping the pressure constant, the α-phase transforms to the denser bcc phase (named β-phase) at about 1163 K. Generally, β-phase Ti is more ductile than the α-phase, due to the larger number of slip planes in the structure of the former in comparison to the latter. β-phase Ti is largely used in biomedical implants due to the match between its mechanical properties and the bones ones [3].
Formation of a hexagonal ω-phase has been reported in high pressure condition (2-7 GPa) [2,4]. Both synthesis and evolution of ω-phase found great attention in the scientific world [5–7] and are object of intense research. Due to its different structure, such phase increases the Young's modulus resulting in an embrittlement of β-phase Ti. Interestingly, some works report about the possibility of inducing the α→ω transition at titanium surface by intense ion or particle beams irradiation in the MeV energy regime [8–11].
In addition to the uses of Ti as bulk material, Ti films are usually employed in a vast range of applications such as in single electron transistors [12], in very-large-scale integration technology and in micro-electro-mechanical systems [13–15]. Moreover, they find application in the field of detection as bolometers in infrared sensors [16], superconducting edge sensors in microcalorimeters [17] and light detectors [18]. Finally, Ti thin films are one of the key components of the targets in laser driven ion acceleration experiments [19].

Considering thin films, it is very difficult to control the crystalline phase using high temperatures, usually not compatible with the substrate, or high pressures. As a result, in most cases, the growing Ti phase is the α-phase. A possible way to control the crystalline phase is to exploit the affinity with the substrate inducing an epitaxial growth. Another strategy involves the addition of a proper ion flux with energies in the hyperthermal range (10s – 100s of eV) [20] to the neutral flux of atoms during the deposition. The use of this technique allows to add another parameter influencing the growth kinetics, morphology and structure of the films. In the hyperthermal energy range, the impinging ions are known to affect the subsurface film layers (1-3 nm). Shallow implantation of bombarding species can cause generation of point defects (i.e. interstitials, vacancies and substitutional atoms). Generally, these defects are unstable and can be annihilated by diffusion towards the nearest under-dense region if sufficient energy is provided, e.g., by an ion field [21].

In the case of Ti, the strategy of epitaxy leads to some results in growing crystalline phases different from the α-phase, but the obtained films thicknesses are in the range of 10s of nm. ω-phase films have been grown pseudo-epitaxially on the surface of Fe (111) by electron beam evaporation in ultra-high vacuum conditions. The transition to the α-phase takes place at approximately 40 nm [22]. A new face centered cubic (fcc) phase, not predicted by the equilibrium phase diagram of Ti, was firstly observed in thin epitaxial films evaporated on NaCl single crystals [23], as well as on metallic [24] and ceramic [25] substrates. In all these works, an fcc→α-phase transition takes place beyond a critical thickness around 1-20 nm.

Respect to the use of energetic species, few examples can be found in literature highlighting the role of an ionic field, mainly argon, on the phase of the Ti films. In the work of Gao et al. [26], the deposition of Ti films by arc-plating with high bias voltage (800 V), at extremely low temperatures (166 K), leads to the coexistence of α and ω phases in the films. The high energy of the species, in addition to the low mobility of the atoms, is addressed as a possible cause of the ω-phase formation. However, the detailed role of such parameter is not discussed.

Chakraborty et al. [27] report about the deposition of a mixture of α and fcc phases in Ti polycrystalline films deposited on Si substrate by Direct Current Magnetron Sputtering (DCMS) with a thickness range of 140 – 720 nm. A thermodynamic stability model is proposed, but the role of process parameters other than film thickness is not discussed. Recently, pure fcc Ti phase films with a thickness of 220 nm have been obtained by cathodic arc discharge by Fazio et al. [28].

To summarize, the deposited Ti coatings usually contain the α-phase only. The possibility to grow Ti films and, at the same time, to tailor the crystalline phase of the growing films in addition to all the other properties would represent a significant step ahead in the use and application of Ti films.

For instance, thin Ti films containing both α and ω phases could exhibit improved mechanical properties like higher strength, hardness and fatigue resistance compared to pure α-phase films. Indeed, this was observed [29] for surface layers containing both α and ω phases in bulk Ti. While the formation of these layers was obtained by means of intense ion bombardment of bulk Ti, the deposition of thin α + ω coatings is still absent in literature. As far as the fcc phase is concerned theoretical studies predict a peculiar electronic structure for fcc Ti that will influence its electronic properties such as conductivity, specific heat and magnetic susceptibility [30].

The use of epitaxy leads to the production of very thin films. The use of energetic species seems to be a promising technique. Thus, it is of extreme interest to address the role of the ion energy environment in determining the synthesis of such phases. In this respect, among the PVD techniques, the High Power Impulse Magnetron Sputtering (HiPIMS) [31] could be an ideal tool to explore such regimes.

HiPIMS relies on the application of very high amplitude voltage pulses [32], separated by long time intervals, to sputter the target cathode. The peak power exceeds the time-averaged power by typically two orders of magnitude [33], leading to a high plasma density [32] ($10^{18}$-$10^{19}$ m$^{-3}$) and thus the generation of metal ions. The energy of such particles can be further controlled by electric fields (i.e. a bias voltage to the substrate), while their flux principally depends on plasma density, target power and magnetic field configuration at the target [31]. By exploiting the features of biased HiPIMS, it is possible to precisely tailor the energy of the ions showing the effect of the energy of the species. It is interesting to note that, differently from many methods that exploit Ar ions as energetic species, in this case the ion and neutral species that contribute to the growth of the films belong to the same element, avoiding the problems related to Ar retention and consequent film embrittlement.

In this work, we investigate how the presence and the energy of the impinging ions determine the evolution of crystalline phases in the Ti growing films comparing DCMS and HiPIMS depositions and varying the deposited thickness. HiPIMS pulse parameters are optimized to maximize the Ti ions flux, and different kinetic energies are chosen. Structure and morphology of the obtained films are characterized by X-Ray Diffraction (XRD) and Scanning Electron Microscopy (SEM). The residual stress level of the growing films is characterized after the deposition by substrate curvature method. We show the possibility of producing Ti films with different crystalline phases. The addition of the Ti ion field allows formation of Ti fcc at low energies and induces the ω-phase formation at high energy in a non-epitaxial regime. We find that the ion density and bias voltage play a crucial role in the fcc and ω phases formation, since they strongly affect the energy and type of film-forming species. In addition we assessed how the energetic ion flux influences the surface mobility of Ti atoms in the 3D Volmer-Weber growth mechanism [34]. Although Ti is classified as a low mobility metal [35], in the HiPIMS deposited films a three-stage compressive-tensile-compressive (CTC) behavior is observed in the development of incremental stress with thickness.

## 2. Experimental details

The Ti films were deposited by HiPIMS and DCMS techniques onto single side 500 μm thick (100) Si wafers and, for stress measurements, onto 300 μm thick double side (100) Si wafers without removal of the native silicon oxide. Prior to film deposition, each Si substrate was cleaned using isopropanol and blown dry in $N_2$. All the depositions were performed in a laboratory-scale high vacuum sputtering system (Kenosistec S.r.l., Italy) equipped with a double pulse generator (Melec GmbH, Germany) used to deposit in DCMS and HiPIMS regimes as well as to bias the substrate. In the vacuum chamber, with a base pressure lower than 5·10$^{-4}$ Pa, a circular titanium target (99.995% purity, Testbourne Ltd, England, UK) with a diameter of 76 mm was placed. To generate the sputtering plasma, the chamber was filled with argon (Ar) as working gas (99.999% purity) with constant inlet gas flow of 80 sccm and pressure equal to 0.5 Pa. The Si wafers were fixed on a top-mounted rotating substrate holder, with a rotational speed of 5 rpm, placed 14 cm away from the sputtered target. In all depositions, the substrate holder was kept at room temperature and no intentional heating was applied.

In the HiPIMS depositions, pulse length, frequency and duty cycle were set equal to 50 μs, 350 Hz and 1.75%, respectively. The pulse voltage ($V_D$) ranged between 400 and 750 V. This parameter was varied to maximize the Ti ion flux (see below in the Results section). The substrate bias ($U_S$), synchronized with the HiPIMS pulse, was applied to modify the energy of the impinging ions. The chosen $U_S$ values were 0, -300 and -500 V. It is worth mentioning that, even if in one case the bias is not applied, the substrate is always subject to a floating potential of about -10 V ÷ -15 V. The waveforms of $V_D$, $U_s$ and discharge current ($I_D$) were characterized exploiting a Rigol DS4034 oscilloscope.

In the DCMS depositions the voltage was varied between 300 and 600 V to obtain powers comparable with the average powers of the HiPIMS depositions. In both regimes, the deposition times ranged between 15 and 180 minutes.

To provide a qualitative comparison between the concentration of the neutral and ionized species, both HiPIMS and DCMS plasmas were analyzed by Optical Emission Spectroscopy (OES) for wavelengths ranging from 365 nm to 865 nm.

Crystalline phase evolution and crystallographic orientation of the films were evaluated by means of XRD analysis performed by a Panalytical X'Pert PRO X-ray diffractometer in $\theta/2\theta$ configuration. The X-ray diffractograms were fitted using the MAUD software [36] to obtain the phase fraction of fcc, ω and α phases of Ti. For the latter phase, we calculated the Texture Coefficient (TC) using the following equation [37]:

$$TC = \frac{I_{(hkl)}/I_{0(hkl)}}{n^{-1}(\sum_0^n I_{(hkl)}/I_{0(hkl)})} \quad [1]$$

where $I_{(hkl)}$ and $I_{0(hkl)}$ are the sample and standard Ti intensities of a peak identified by Miller indices hkl, while $n$ is the number of considered peaks.

Morphological properties were assessed by a Zeiss Supra 40 field emission Scanning Electron Microscope (SEM), operating at an accelerating voltage of 5 kV. The film thicknesses were determined by SEM cross-section images. After the cleavage of the samples at least 4 images were taken, and the average thickness was evaluated. Thickness evaluation is affected by the film morphology resulting in a maximum uncertainty of about 7-10%.

Residual stresses were measured by an optical implementation of the wafer curvature method. The ad-hoc developed experimental set-up, already described and used in previous works [38,39], exploits a set of parallel laser beams to probe the curvature radius of the coating-substrate system. The laser beams impinge on the uncoated substrate surface and then they are collected by a high frame rate camera. Residual stresses were quantified by measuring the curvature variation of the uncoated surface of the double side Si wafer before and after the film deposition. According to the Stoney equation [40], the residual stress $\sigma_{res}$ can be calculated as:

$$\sigma_{res} = \frac{E_s}{1-\nu_s} \cdot \frac{t_s^2}{t_f} \cdot \frac{1}{6R_c} \quad [2]$$

where $E_s$ and $\nu_s$ are the Young's modulus and the Poisson's ratio of the substrate, $t_s$ and $t_f$ are the thicknesses of the substrate and of the film and $R_c$ is the curvature radio of the system. To properly quantify $R_c$, we performed multiple measurements on the same sample varying the probed position.

## 3. Results

In the first part of our investigation, we will describe the process parameters related to the optimization of the HiPIMS plasma with the aim of maximizing the $Ti^+/Ti$ ratio and, thus, the effects related to the ion bombardment during the growth of Ti films. Secondly, we will present the characterization of Ti films deposited in a low and high ion fraction environment exploiting DCMS and HiPIMS regimes, respectively. We will refer to the first as Ti-DCMS films and to the second as Ti-HiPIMS films. The latter were deposited both in a low ion energy ($U_s$ = 0 V) and high ion energy ($U_S$= -300 V and -500 V) environment. The deposition strategy, aimed at investigating the influence of both type and energy of film forming species on film characteristics, is summarized in Table 1.

Table 1. Summary of the adopted deposition strategy. DCMS regime allows to investigate films grown in a low ion fraction and low ion energy environment. Unbiased HiPIMS regime allows to investigate films grown in a high ion fraction and low ion energy environment. Biased HiPIMS regime allows to investigate films grown in a high ion fraction and high ion energy environment.

| Target metal ion fraction | Energy of the ions | Explored thicknesses |
|---|---|---|
| Low (DCMS) | Unbiased | 80 - 650 nm |
| High (HiPIMS) | Unbiased | 50 – 600 nm |
| | 300 V | 30 – 600 nm |
| | 500 V | 30 – 500 nm |

### 3.1 HiPIMS I-V characteristic

Fig. 1a shows the HiPIMS discharge pulse waveforms. The discharge voltage waveform exhibits a rectangular shape (i.e. the voltage is approximately constant during the pulse on-time), while the discharge current increases until the end of the pulse with the typical peak shape.

Fig. 1b presents the discharge peak current ($I_{D, peak}$) vs discharge voltage ($V_D$) characteristic in a semi-log plot varying $V_D$ between 400 and 750 V. As reported in literature [41,42], an exponential relationship $I_{D,Peak} \propto V_D^n$ exists between the target voltage and the peak current.

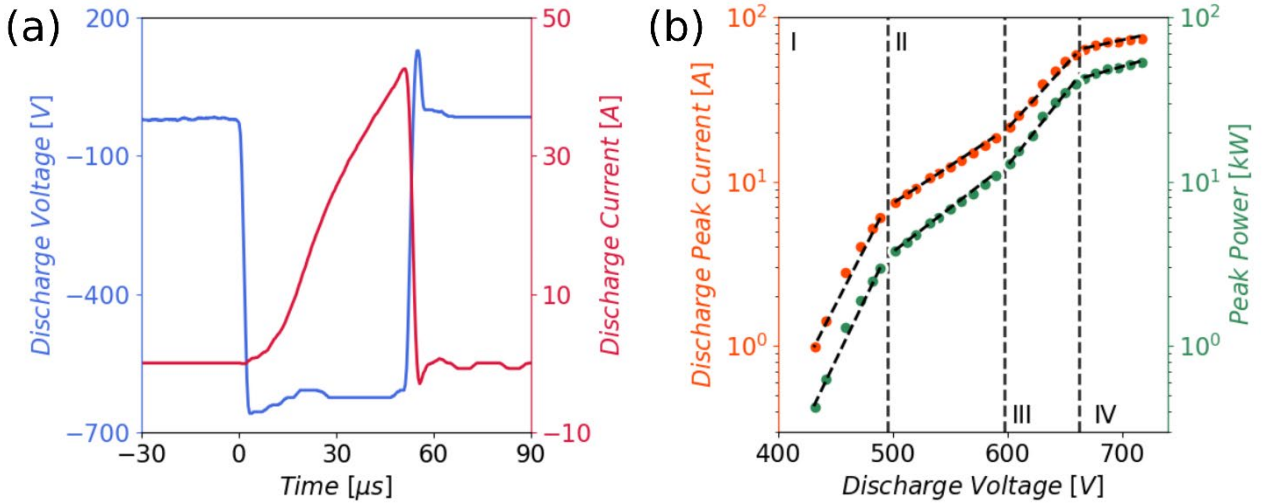

Fig. 1. (a) Experimental voltage and current waveforms taken at 0.5 Pa Ar pressure using 50 μs long HiPIMS discharge pulses on a Ti target. (b) Experimental trends of discharge peak current (orange points) and peak power (green points) vs discharge voltage ($V_D$) taken at 0.5 Pa Ar pressure. The vertical dashed black lines highlight the four regions with different $n$ value.

The value $n$ changes increasing $V_D$, independently from the pulse on/off-times, and accordingly four distinct regions (I, II, III, IV) can be recognized [31].

We varied $I_{D, Peak}$ and $V_D$ between 10-55 A and 550-660 V, respectively. Thus, we worked in the voltage range which approximately corresponds to the III region, characterized by a rapid growth of $I_{D, Peak}$. However, the I-V curve is affected by the cathode geometry. Thus, to present the trend in a more general way according to a recent work of Wu et al. [43], we used the peak power $P_{Peak}$ defined as $P_{Peak} = I_{D, Peak} \cdot V_D$ instead of the parameter $I_{D, Peak}$. Indeed, it has been shown that this parameter plays a key role in determining the ion to atom ratio [44,45]. The trend of $P_{Peak}$ against $V_D$ is also reported in Fig. 1b.

*3.2 Plasma characterization with OES*

Fig. 2 displays optical emission spectra taken in DCMS (blue) and in HiPIMS (red) regimes at the same power (average power in the HiPIMS case) of approximately 450 W. In both spectra, two different regions of peaks are present. In the first region, roughly from 365 nm to 550 nm, most of the lines refer to titanium species, both neutrals and ions, while a few lines only refer to ionized argon species. In the second region, roughly from 680 nm to 865 nm, the lines refer to argon species, mainly the neutral ones. By comparing the spectra recorded in the two regimes, the HiPIMS spectrum shows a significant increment in titanium species emission with respect to the DCMS case, in which there is a greater contribution from the argon species.

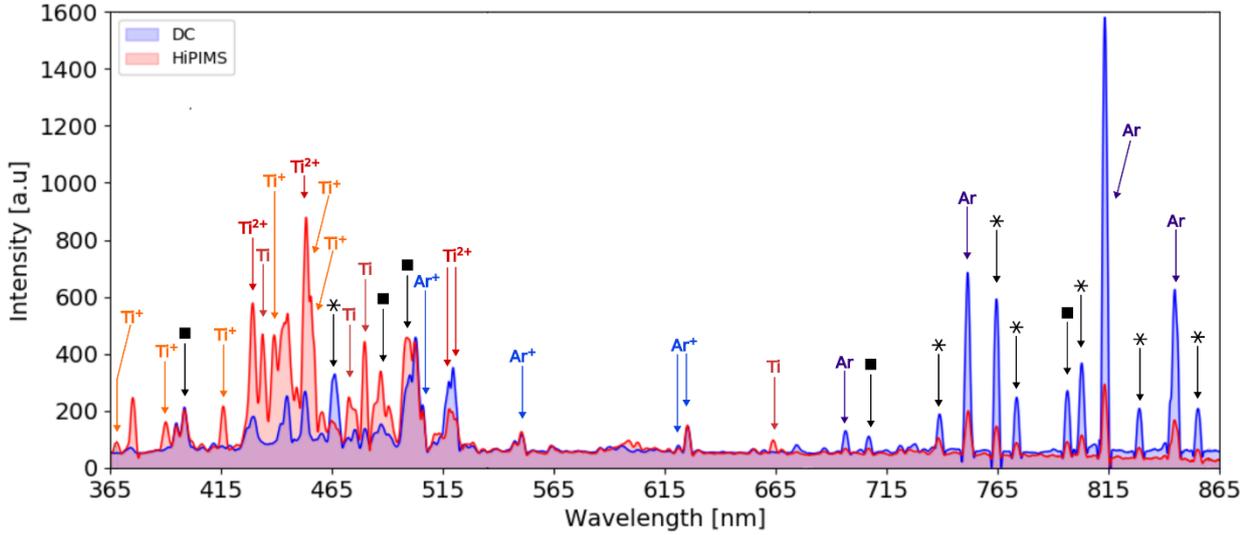

**Fig. 2.** Optical emission spectra acquired in DCMS (blue) and HiPIMS (red) regimes at the same power (average power in the HiPIMS case) of about 450 W. Colored arrows indicate the emission lines for Ti, $Ti^+$, $Ti^{2+}$, Ar and $Ar^+$ identified with the procedure described in Appedix A and used to evaluate the emission intensity ratios for the HiPIMS regime. Peaks identified by an asterisk belong to Ar or $Ar^+$ species, while peaks identified by a black square belong to Ti, $Ti^+$ or $Ti^{2+}$ species, accordingly only to the NIST database.

We studied the effect of $P_{Peak}$ on the plasma composition. To this aim, we varied $P_{Peak}$ in the range 5-50 kW and we performed OES measurements. Fig. 3 shows the emission intensity ratios $I(Ti^+)/I(Ti)$ and $I(Ar^+)/I(Ar)$ for the HiPIMS regime as a function of the $P_{Peak}$. The intensities $I$ were evaluated as the average intensity performed over specific peaks of the OES spectrum. The procedure employed to calculate the average intensities, as well as the procedure adopted for the peak identification, is reported in Appendix A. In particular, the selected emission lines of Ti and $Ti^+$ exhibit similar excitation energies and this allows to exclude the dependence of excitation probability on excitation energy [43,45]. In case of Ar and $Ar^+$, the difference in the excitation energies is more marked. However, as a first approximation, the same reasoning of Ti species applies since the energies are of the same order of magnitude. On the other hand, we excluded $Ti^{2+}$ species from this analysis since their excitation energies are too high compared to the Ti and $Ti^+$ ones. The emission intensity ratios show a maximum value in the region between 20 and 30 kW, where the number of ionized species is maximized with respect to the number of neutral species. Thus, a peak power of 27 kW has been chosen for all the HiPIMS depositions. Having maximized the Ti ion flux, the deposition of Ti films with different thickness (setting the deposition time) and energy of the ions (setting the bias) has been performed. In addition, some vacuum annealing procedures have been made. A complete list of the deposited samples, treatments and characterizations is reported in Appendix B.

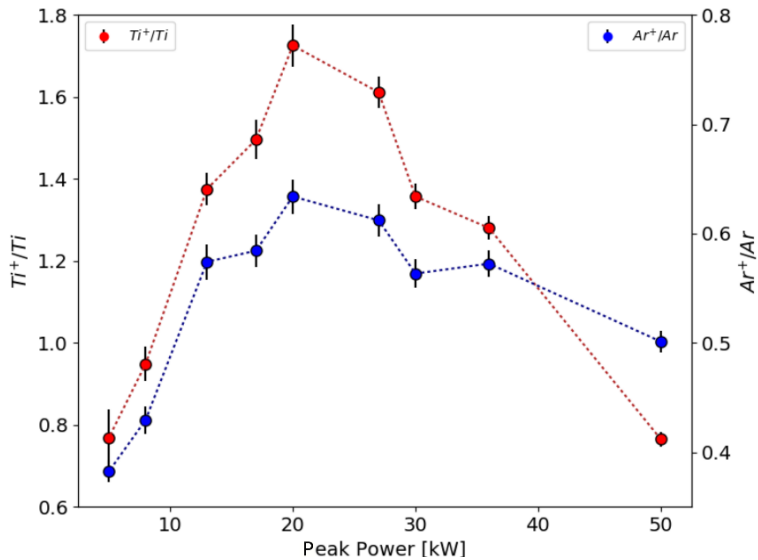

**Fig. 3.** Emission intensity ratios calculated for the average intensities of the peaks $Ti^+/Ti$ and $Ar^+/Ar$ as a function of $P_{Peak}$.

## 3.3 Phase determination and oriented growth characterization with XRD

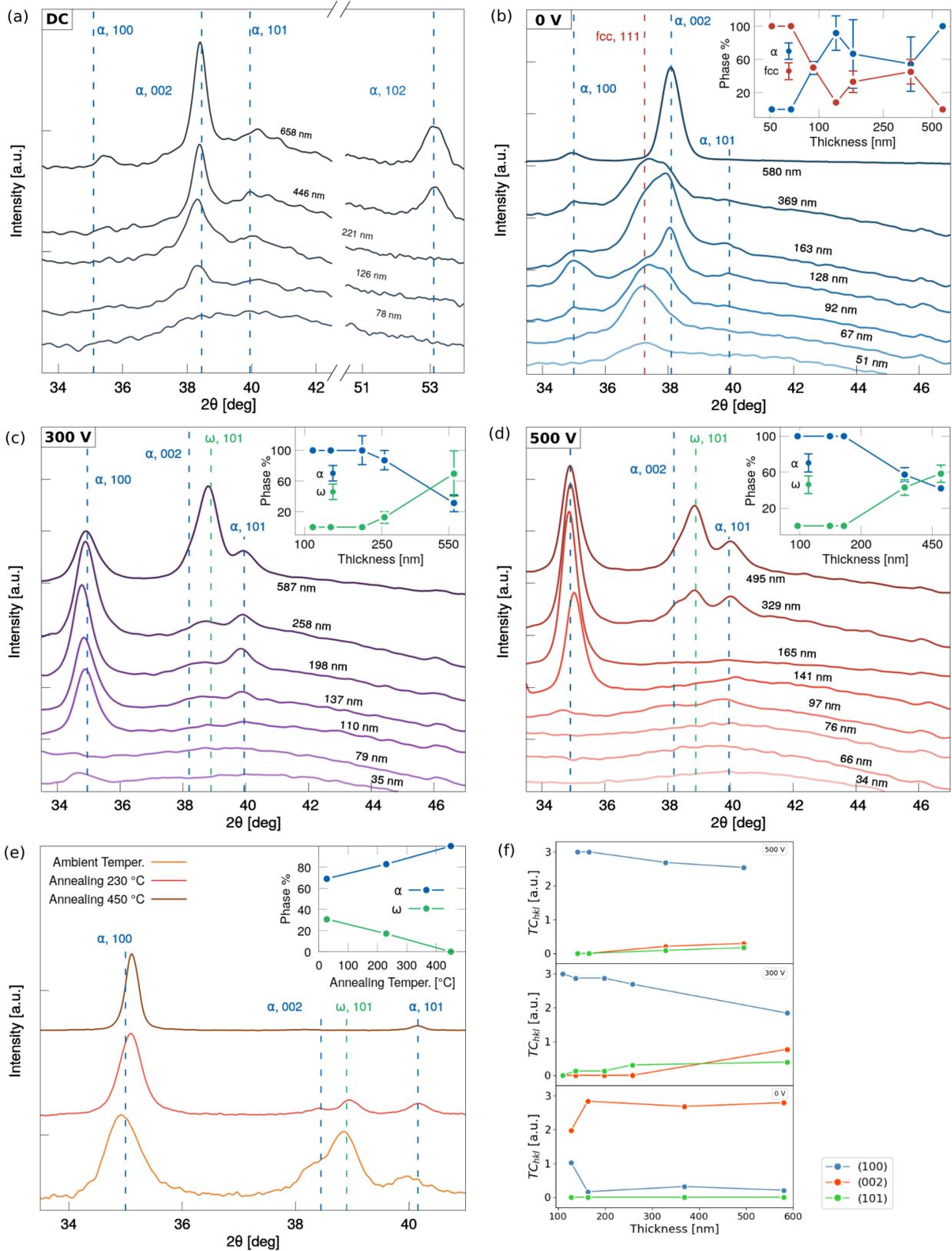

**Fig. 4.** XRD spectra of Ti films deposited in (a) DCMS regime, (b) HiPIMS regime without an applied bias, (c) HiPIMS regime with an applied bias of 300 V. (d) HiPIMS regime with an applied bias of 500 V. For figure (a)-(d) the corresponding films thickness are reported above the spectra. (e) XRD spectra of a Ti films before and after annealing procedure. Insets in figure (a)-(e) report the phase fraction variation as function of annealing temperature. (f) Texture coefficient of α-phase films at different bias.

To investigate the phase evolution in Ti films, XRD spectra of films grown under different deposition conditions were acquired. Fig. 4 presents the results of this analysis.

Fig. 4a shows the XRD spectra of Ti-DCMS films, for thicknesses ranging approximately in the interval 80-650 nm. As far as the thinnest film is concerned, no peaks can be identified in the XRD spectrum. In the other spectra we recognize peaks related to the α-phase of Ti only, as reported in many previous works [46–48]. For intermediate thicknesses (i.e. roughly 100-200 nm) we observe only the (002) reflection at $2\theta = 38.41°$, while (100), (101) and (102) reflections (respectively at $2\theta = 35.07°$, $40.153°$ and $53.012°$) emerged for increasing deposition times.

Fig. 4b shows the XRD spectra of Ti-HiPIMS films without an applied bias, for thicknesses ranging in the interval 50-600 nm approximately. In the 50 nm thick film only one peak positioned at $2\theta = 37.05°$ is visible. According to previous works [27,28], we associate the peak to the Ti fcc phase and we identify it as the (111) reflection. Above a thickness of 100 nm, the XRD spectra reveal the presence also of two α-phase reflections, the (100) and the (002). For a relatively wide range of thicknesses, we have the coexistence between the two phases. As shown in the inset of Fig. 4b, the α-phase gradually becomes predominant as the thickness increases. Since in the last spectrum of Fig. 4b only peaks associated to α-phase are present, we assumed a complete fcc→α-phase transition at a critical film thickness between 370 nm and 580 nm. It is worth to note that, after the complete transformation of the fcc in the α-phase, the resulting (002) peak is significantly shifted with respect to its original position, probably due to residual stress. Specifically, the shift is of about 0.3° respect to the (002) reflection of the Ti-DCMS films. The inset in Fig. 4b reports the fcc and α-phase fractions as a function of sample thickness evaluated with the MAUD software.

Fig. 4c and Fig. 4d show the XRD spectra of Ti-HiPIMS films deposited in presence of an applied negative bias of 300 V and 500 V, respectively. The film thicknesses vary in the interval ~ 35-600 nm approximately. In both cases, up to a thickness of 100 nm, the spectra did not exhibit the presence of peaks. As the thickness further increases, α-phase peaks can be recognized in the diffractograms. The growing films appear to be strongly oriented along the (100) direction. In addition, we observed a peak corresponding to the (101) orientation. At a thickness of about 300 nm a new peak appears at $2\theta = 39°$. We associated this signal to the (101) reflection of the ω-phase which, according to literature [49], is located at $2\theta = 39.01°$. The insets in Fig. 4c and Fig. 4d report the ω and α-phase fractions as a function of samples thickness evaluated with the MAUD software.

Fig. 4e reports the XRD spectra associated to Ti-HiPIMS films before and after an annealing procedure. Films characterized by the presence of the ω-phase (101) peak were annealed at two different temperature of 230 °C and 450 °C. The spectra reveal a decrease up to a complete absence of the ω-phase peak.

Lastly, we report in Fig. 4f the texture coefficient evolution of the Ti α-phase for the unbiased and biased depositions. At 0 V a prevalence of the (002) orientation is clearly visible, even if we still observe the (100) signal. On the contrary, in presence of a 300 V and 500 V bias, the (100) orientation is the preferred one although both the (002) and (101) peaks appear at a certain thickness.

## 3.4 Morphological characterization with SEM

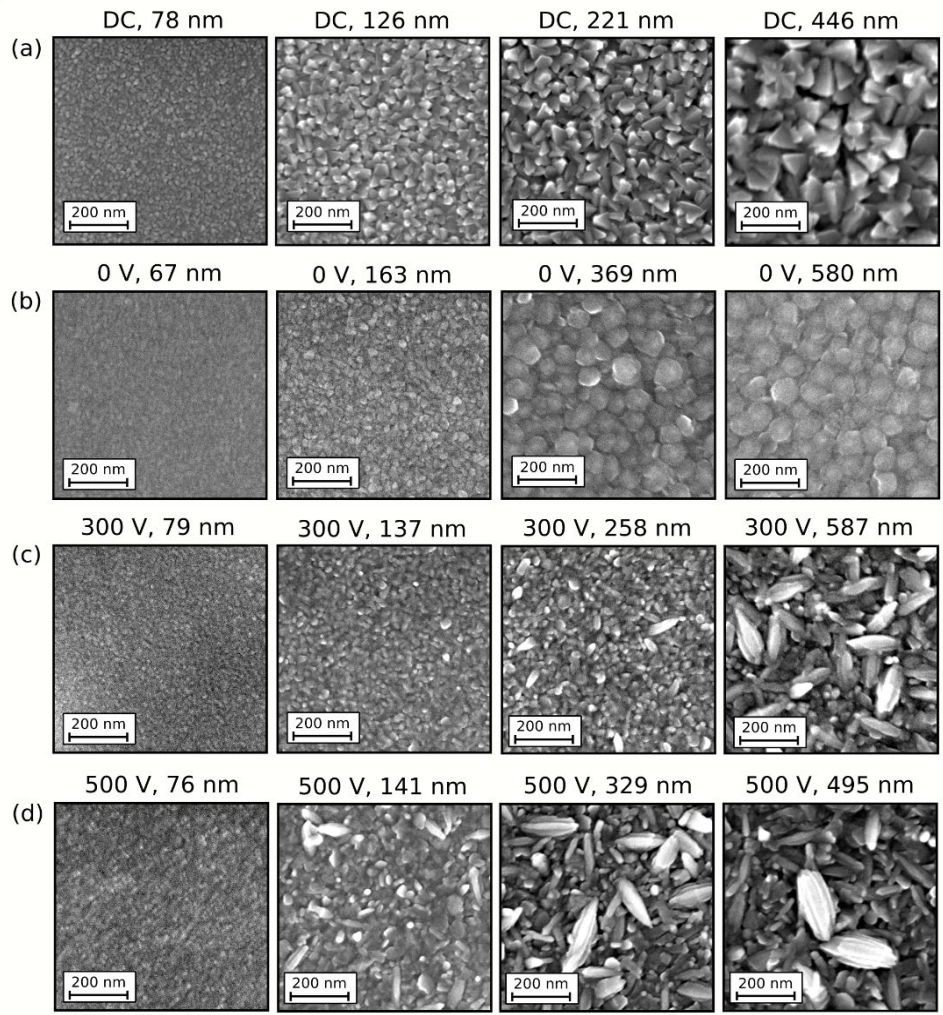

**Fig. 5.** SEM plane view images of Ti films with different thicknesses deposited in (a) DCMS regime, (b) HiPIMS regime without an applied bias, (c) HiPIMS regime with an applied bias of 300 V, (d) HiPIMS regime with an applied bias of 500 V.

As displayed by SEM plane view images in Fig. 5, the various films deposited in DCMS and HiPIMS regimes show different morphologies. Ti-DCMS films (Fig. 5a) exhibit a columnar structure with a pyramidal shaped superficial pattern. The observed structure corresponds to zone I in the structure-zone diagram proposed by Thornton [50] and extended for energetic depositions by Anders [51]. The columnar morphology is confirmed also in the cross-section SEM image displayed in Fig. 6a.

As far as the 0 V bias voltage is concerned (Fig. 5b), we distinguish a granular morphology for all the thickness range considered. According to the XRD analysis, the grains of the thinnest films are associated to the fcc phase, while at higher thicknesses the hexagonal grains belonging to the α-phase appear. Cross sectional image (Fig. 6b) shows a much more compact morphology composed of nanometric grains and very different with respect to that of DCMS case, highlighting the features of HiPIMS deposition even at low energy. Fig. 5c and Fig. 5d display the evolution of the Ti-HiPIMS films surfaces for an applied bias voltage of 300 V and 500 V, respectively. In both cases, for thicknesses greater than 100 nm, we observe the formation of elongated lamellar grains. SEM cross sections (visible in Fig. 6c and Fig. 6d) show different features with respect to the previous cases. In these energy conditions, nanovoids are visible and the section appears as the result of a plastic deformation. The quantity and dimension of these grains increase as the films become thicker. Moreover, a higher value of the applied bias voltage seems to anticipate the nucleation of these structures. In Fig. 7, more details related to these surface structures can be appreciated. As visible in the tilted SEM image (Fig. 7a), it is possible to see that the lamellar grains protrude from film surface of about 50 – 100 nm. Furthermore, the high magnification image of Fig. 7b shows that such grains are constituted of aligned structures that develop in form of "petals".

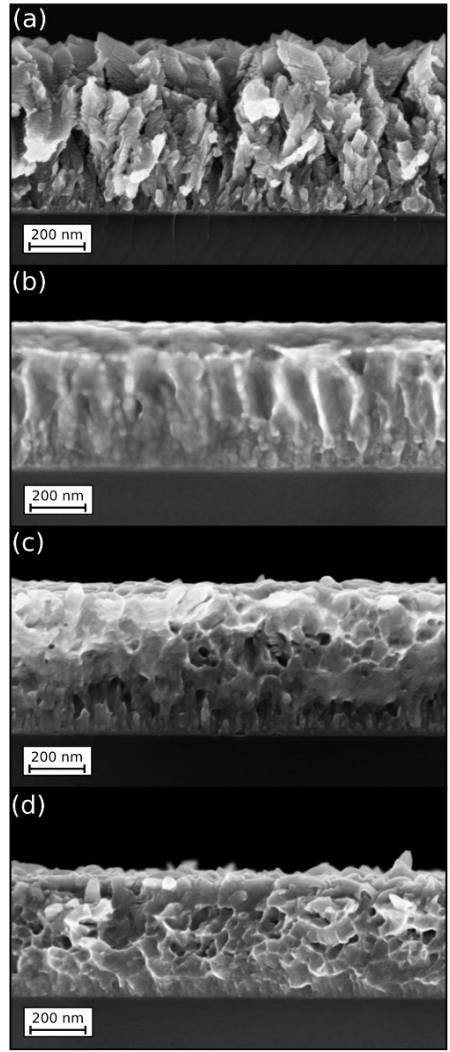

**Fig. 6.** SEM cross section images of thickest Ti films deposited in (a) DCMS regime, (b) HiPIMS regime without an applied bias, (c) HiPIMS regime with an applied bias of 300 V, (d) HiPIMS regime with an applied bias of 500 V.

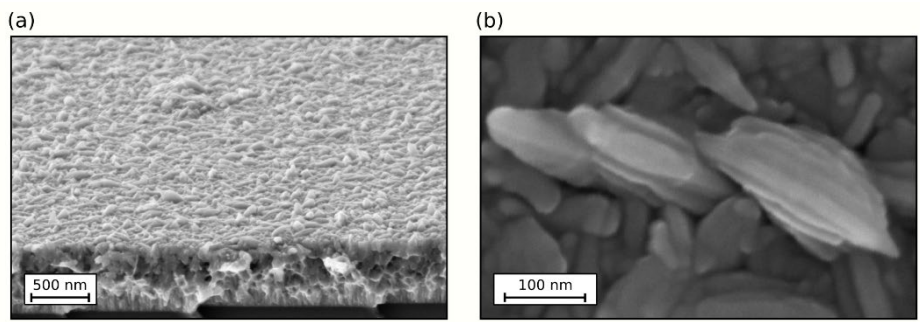

**Fig. 7.** (a) 20° tilted cross section view of the thickest HiPIMS film deposited with an applied bias of 500 V. (b) Magnification image of some lamellar grains.

*3.5 Stress analysis by substrate curvature method*

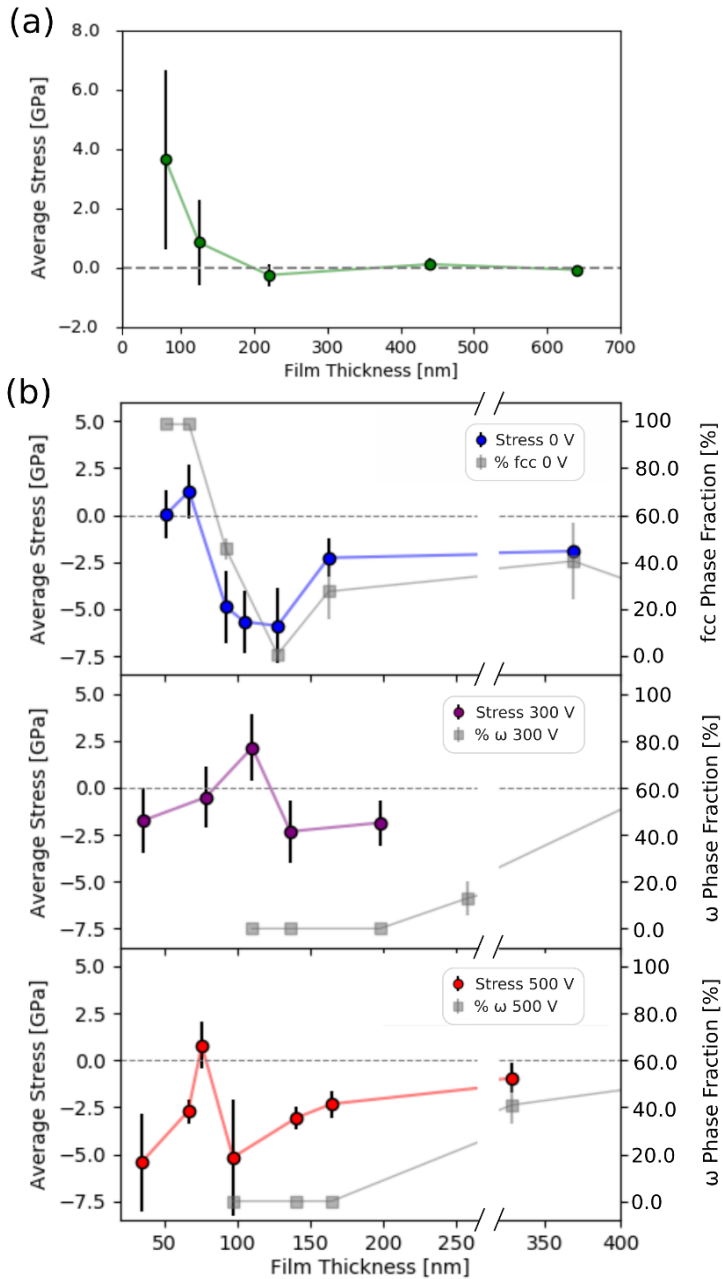

**Fig. 8.** (a) Average residual stress as a function of thickness for Ti films deposited in DCMS regime. (b) Panel 1: Average residual stress as a function of thickness for Ti films deposited in HiPIMS regime without an applied bias. The fcc phase fraction variation as a function of films thickness is plotted. Panel 2 and 3: Average residual stress as a function of thickness for Ti films deposited in HiPIMS regime with an applied bias of 300 V and 500 V, respectively. The ω-phase fraction variation as a function of films thickness is plotted.

The wafer curvature method returns the average residual stress of the Ti films deposited on the Si substrate. Despite the samples could be characterized by a non-uniform stress state, the average residual stress can be helpful to obtain trends varying the deposition parameters. The average residual stresses, as a function of films thickness and bias voltage, are reported in Fig. 8a and Fig. 8b for Ti-DCMS and Ti-HiPIMS films, respectively.

The Ti-DCMS average stress trend (Fig. 8a) shows an initial high tensile state (about 4 GPa), then a stress relaxation occurs for increasing thickness. For films thicker than 200 nm the average stress state is almost negligible.

On the contrary, the average stresses associated to the HiPIMS depositions (Fig. 8b) are characterized by a completely different behavior that, in addition, is affected by the energy of the ion species. At low ion energy (10-15 eV, $U_s$ = 0 V) in the first growth stages (50 nm thick) the stress state is almost zero then it becomes tensile (1 GPa). However, an increment in the deposited thickness rapidly induces a highly compressive stress state and it reaches a value lower than -5 GPa at 128 nm. A further increasing of films thickness induces a stabilization of the stress at about -2.5 GPa. At high ion energies (100s eV, $U_s$ = 300 V and 500 V), for low thicknesses a highly compressive state (about -2 and -5 GPa, respectively) is found. Then, increasing films thickness, the state becomes tensile (2 and 1 GPa) and, finally, it becomes again compressive with a sort of relaxation.

## 4. Discussion

Over the last two decades, many efforts have been made to develop deposition techniques which allow a better control of the energy and flux of the involved species, since they greatly affect both microstructure and phase composition of the growing films [20]. Thanks to its high-density, the HiPIMS plasma is an ideal tool to produce an energy flux in the hyperthermal range (10-1000 eV), using a proper substrate bias voltage. The phase tailoring of tantalum [52] and $TiO_x$ [53–55] films have been achieved by properly setting the energy of the impinging ions in the range of 10 – 90 eV and few eV, respectively. The formation of a specific phase is influenced by the energies of the particles striking the substrate, but also by the type of incident species, both those of the working gas and the metal ones [52,53]. In addition, also the contribution of the substrate can be "activated" depending on the energy of the species. In the works of Cemin et al. [35,56], a significant change in the crystallographic orientation of Cu films from the (111) to the (002) direction has been related to the energy of the ions that are able to cross the native Si oxide barrier and induce the epitaxial growth of Cu onto (100) Si substrate. To estimate the penetration depth of $Ti^+$ ions in the native silicon oxide layer covering the substrate we performed SRIM simulations [57]. We estimated a thickness of the native Si oxide of about 0.7 – 1.5 nm [58] and we selected the energies of the ions ranging from 20 eV (lower limit energy) to 500 eV. The results are displayed in Table 2.

**Table 2.** Range of Ti ions with five different energies in Si oxide layers over semi-infinite Si substrates and in semi-infinite Ti. The average ranges and uncertainty (i.e. longitudinal straggling) are evaluated with the SRIM code.

| Energy | 20 eV | 30 eV | 60 eV | 300 eV | 500 eV |
|---|---|---|---|---|---|
| $SiO_2$ (0.7 nm) + Si (inf.) | 0.8 ± 0.1 nm | 0.9 ± 0.2 nm | 1.1 ± 0.2 nm | 2.0 ± 0.7 nm | 2.5 ± 0.9 nm |
| $SiO_2$ (1.5 nm) + Si (inf.) | 0.9 ± 0.1 nm | 1.0 ± 0.2 nm | 1.2 ± 0.3 nm | 2.2 ± 0.6 nm | 2.7 ± 0.8 nm |
| Ti (inf.) | 0.5 ± 0.1 nm | 0.6 ± 0.2 nm | 0.7 ± 0.2 nm | 1.2 ± 0.6 nm | 1.6 ± 0.8 nm |

It can be noted that an energy of 20 – 30 eV leads to penetration depths lower than the thickness of the native Si oxide, whereas energies of 300 and 500 eV largely overcome the thickness of the native oxide. Thus, we decided to investigate the effect of the ion energy considering two distinct ranges. A low energy range ($U_s$ = 0 V, energy of the $Ti^+$ species of 10-15 eV) where the interaction with the underlying (100) Si crystal is negligible, and a high energy range ($U_s$ = 300 V and 500 V) where we have coupling with the Si substrate as well as intermixing and sputtering of the Si oxide. Moreover, in the second energy regime, a high compressive stress state can be induced by defect formation, atomic-peening or grain boundary densification [59]. It is worth mentioning that the stress evolution during the films deposition and the microstructural development of the films are strongly related [34,60].

As a reference, we performed the deposition of Ti-DCMS films without bias and with the same average power of the HiPIMS depositions. This regime is characterized by the presence of species (mainly Ar and Ti neutrals) with energies of 1-3 eV. As widely reported in literature [46–48,61,62], this kind of energy regime results in the growth of the α-phase with preferred (002) orientation due to the minimization of the surface energy [63].

Moreover, increasing films thickness, a tensile stress state develops and this is the typical trend observed in low mobility metals characterized by a 3D Volmer-Weber growth [34]. The basic parameter affecting mobility is the so-called homologous temperature $T_h = T_s/T_m$, where $T_s$ is the substrate temperature, while $T_m$ is the melting temperature of the deposited material. There is a widespread agreement that films deposited at $T_h<0.2$ results in a low mobility, while for $T_h>0.2$ the conditions for high atomic mobility are reached. In our case, under the assumption of a substrate at room temperature, $T_h$ results 0.15 and therefore we should expect a tensile trend similar to what observed for Ti-DCMS films.

Considering the corresponding Ti-HiPIMS films ($U_s= 0$ V) from the presented results, it can be clearly seen that Ti depositions, where in addition to the atomic flux we have presence of a low energy ion flux, result in the nucleation of the fcc phase confirming the ability of ions in nucleating this kind of phase. Because of the low duty cycle of the deposition technique and the low energy of the impinging ions in this regime, we do not expect a significant variation of the substrate temperature during the film deposition and, thus, an effect of temperature on the features of the film.

As already stated, we selected this energy regime with the aim of excluding any interaction with the Si surface. Thus, although we cannot completely exclude a contribution coming from the substrate, we hypothesize that other mechanisms can play a more important role in the nucleation of the fcc phase. Indeed, we can have some hints about the nucleation dynamics by analyzing the stress state development. As already mentioned, the stress evolution of Ti-DCMS films in the present deposition conditions resembles a low mobility metal [64]. However, the stress state behavior found in Ti-HiPIMS films increasing thickness shows the compressive-tensile-compressive (CTC) behavior [35]. This is usually ascribable to high surface mobility metals ($T_h>0.2$) following the Volmer-Weber growth mechanism involving nucleation of 3D islands and the subsequent growth, impingement and coalescence of the same to form continuous films [34]. As reported by Magnfält et al.[21], when low mobility metals (as Ti) are deposited under a large flux of hyperthermal species, whose energy range affects the first layers of the film, the migration of the adatoms at the under-dense grain boundaries can be achieved. Evidently in the present energy regime, due to the contribution of the migrating adatoms, the nucleation of the fcc phase is favored with respect to the α-phase at least a low thickness. In this respect, it is interesting to note that the signal coming from the fcc phase is detectable already for 50 nm thick films, whereas the α-phase in Ti-DCMS can be seen starting from 120 nm.

It is interesting to compare our findings with the few results present in literature where an ion field is used. The first indications of the role of low energy ions and atoms in nucleating the fcc phase were reported by Yue et al. who deposited Ti films by ion-beam sputtering [65]. Moreover, the ion energy regime present in cathodic arc deposition is sufficient to induce the nucleation of the fcc phase as reported in the work of Fazio et al. [28]. From the thermodynamical point of view, they claim the generation of the fcc phase to be related to the minimization of the interfacial energy with the substrate.

Although not present in the phase diagram, a local stability of the fcc phase in Ti, Hf and Zr has been theoretically demonstrated [30]. In that work, the total energy of the fcc phase resulted higher than the hcp and ω phases, but lower than the bcc phase. Thus, it is possible to image that the presence of an out of equilibrium situation, as the growth in form of film, in addition with a suitable enhanced "mobility" of the impinging atoms can induce the fcc growth instead of the usual, and more thermodynamically stable, hcp phase.

Increasing the deposition time, we have found a transition from the fcc to the α-phase. This transition has been found in all the investigated Ti fcc films. In the work of Chakraborty et al. [27], a thermodynamic model is presented to address the fcc → hcp transition by means of bulk, surface and strain free energy change. The increased stability of the hcp phase is related to the bulk free energy change that grows increasing film thickness. In that model a critical thickness of 35 nm is foreseen for the stability of the fcc phase. However, in literature it is reported that the transition thickness greatly depends on the deposition conditions ranging from few to hundred nanometers. In our case we found pure fcc phase up to a thickness of 67 nm and a mixture of fcc and hcp phases in the range 92 - 370 nm.

Moreover, it is worth highlighting the connection between fcc phase content and the stress state. In Fig. 6b, the fcc phase concentration is plotted as a function of the thickness. A strong correlation between the fcc phase fraction and the average stress state can be observed. In this energy regime, the mobility of the adatoms is possible but is not high. In addition, the implantation of Ti ions results in the formation of interstitial defects. The combination of these effects, increasing the deposition time, results in the development of high compressive stresses also as a result of grain boundaries densification [21,35]. As reported in recent theoretical works [66,67] related to the evolution of Ti fcc and α phases under compressive and tensile loadings, the presence of a compressive state induces the formation of Schokley partial dislocations that slip the crystalline planes from the (111) direction of the fcc to the (002) direction of the Ti α-phase. Interestingly in [66], a critical value of 2.5 GPa is calculated for the starting of the transition. Our observations agree with this value, since at 2.5 GPa we have a sensible reduction of the fcc phase. Thus, the ion field seems to produce a double effect. On one side, it activates the mobility of the adatoms inducing the nucleation of the fcc phase. On the other side, it introduces defects leading to an increase in the compressive stress state that, in turn, promotes the formation of the α-phase. A fine tuning of the energy could further increase the maximum thickness for fcc containing films. Lastly, we note that the synthesis of the fcc phase is related to the use of low energy species. Indeed, the films grown in high energy regime do not promote the formation of the fcc phase.

In the high energy regime, for a thickness lower than about 100 nm, no phase identification is possible. Starting from 100 nm the (100) is the only visible orientation of α-phase. The development of such orientation can be ascribed to the strain state of the system. In fact, as discussed by Checchetto [63] in terms of minimization of surface and bulk strain energy, the (002) orientation of α-phase is related to the minimization of the surface energy when the strain contribution is negligible, while the (100) orientation minimizes the strain energy of the system. In our case, the presence of a high energy field of impinging particles results in an increment of the strain energy of the system and, in turn, in the development of the (100) direction instead of the (002). It is interesting to note that in agreement with our considerations, the α-phase Ti films deposited by Chakraborty [27] in a low energy environment (standard DCMS) exhibit the (002) orientation. On the other hand, the α-phase Ti films deposited in the high energy environment of cathodic arc [28] shown the (100) growth direction.

Looking at the stress state development, it is easy to notice that, especially in the first layers, the high energy ion field induces higher adhesion of Ti islands onto the substrate, due to atom intermixing and sputter etching of the native Si, determining a strong compressive stress in the system. Then, after reaching a tensile peak associated to islands impingement and coalescence, trends show an increment in compressive stress associated to the continuous film growth. A maximum compression is reached, after which a relaxation occurs as the films thickness further increases. During this high stress tensile-to-compressive transition only α-phase is present, as visible in Fig. 6b.

When thickness is about 300 nm and the stress state of the system is relatively low (1-2 GPa in compression), both for 300 V and 500 V deposited Ti films, we see the development of the (101) peak of the ω-phase.

Although the development of the films grown at 300 V and 500 V are similar there are some remarkable differences. It is interesting to note that a shift of the CTC peak at higher thickness is visible in samples deposited at 300 V with respect to the unbiased situation. This is a clear indication of the increased mobility of the adatoms on the Si surface [68]. On the other side, in the samples deposited at 500 V there is no shift of the CTC peak with respect to the unbiased situation. In this case, the effect of the Ti ions seems to be more related to the production of defects and substrate sputtering than to the enhancement of adatom mobility. Indeed, 500 V Ti-HiPIMS films are more stressed and lead to an increase of the texture coefficient (see Fig. 4f) of the (100) reflection that, as already stated, minimizes the strain energy of the system.

The temperature stability of the ω-phase has been characterized by vacuum annealing. As reported in literature [6], the ω→α transition takes place during an annealing procedure at a temperature which depends on the heating rate. In accordance with the above-mentioned ω→α transition, we have observed a partial transformation at 230 °C and a complete transformation at 450 °C. This is a further evidence of the ω-phase presence in the thickest films deposited in presence of a substrate bias voltage. Moreover, it is interesting to notice that in our case the development of such phase is correlated with the growth of large grains in the films with a stratified morphology, which remain present even after annealing.

However, the ω-phase nucleation in biased Ti-HiPIMS films seems quite unusual. The formation of ω-phase in bulk is associated to high stress state and all the studies show a connection with high pressure, while in the biased Ti-HiPIMS films ω-phase would seem to appear during compressive stress relaxation. As reported in Table 2, in the high energy regime the range of Ti ions in the already deposited Ti film is of about 1.5 nm (i.e. many atomic layers). This implies high mobility of the Ti species, leading to the complex CTC behavior plotted in Fig. 6b, but also defects formation. This out of equilibrium condition could be the mechanism originating the evolution of one of the α-phase orientations that, in the final unloading step of CTC transition, finds a new energy minimum by rearranging itself to the structure of ω-phase. Due to the intense ion energy and flux, especially at 500 V, we cannot exclude an increment of the growth temperature. As shown in the annealing experiments, ω → α transition is likely to occur already at 230°C. Thus at 500 V, ω-phase formation could be hindered also by the slight increase of substrate temperature.

## 5. Conclusion

We have shown that, thanks to the use of HiPIMS technique, it is possible to obtain Ti films exhibiting not only the α-phase, but also other uncommon crystalline phases. With respect to conventional DCMS, HiPIMS allows to properly tailor the energy of the ions and their density in the plasma which, for the purposes of this work, has been maximized by analyzing the plasma composition in a range of $P_{Peak}$ from 5 to 50 kW. In this way, we were able to grow and compare Ti films both in a low Ti ions (DCMS) and high Ti ions (HiPIMS) environments. Furthermore, the application of different substrate bias voltages during HiPIMS depositions allows us to investigate both low and high energy regimes for Ti ions. While the DCMS samples show the presence of α-phase only, the HiPIMS samples contain the unusual fcc and the ω phases.

Here we summarize the most relevant results:

a) The fcc phase is formed in a low energy ion environment (10 – 15 eV). Its evolution is related to the stress state of the system that induces the transition to the α-phase, oriented along the (002) direction, probably via Shockley partial dislocation slip.
b) In presence of a high energy environment (300 – 500 eV), the formation of α-phase oriented along the (100) direction has been obtained as a result of the minimization of the strain state of the system induced by the high energy impinging ions.
c) At "high" film thicknesses the high energy regime drives to ω-phase nucleation. Contrary to Ti bulk and epitaxial growth under bombardment of high energy ions, the formation of ω-phase seems to take place in thick films with relatively low amount of stress (about 1-2 GPa in compression). This leads to the coexistence of α and ω phases that appears to be favored with respect to a complete reverse transformation to α-phase.

The production of Ti films, with different thicknesses, for which tuning the phase composition is possible, could be of interest for various existing industrial and engineering applications to improve the performances of Ti films. Moreover, being able to properly tailor Ti films phase composition can be of paramount importance even in innovative applications, as the production of engineered targets for laser driven ion acceleration experiments, where there are peculiar and precise requirements on the final properties of the Ti films.

## Acknowledgements

This project has received funding from the European Research Council (ERC) under the European Union's Horizon 2020 research and innovation programme (ENSURE grant agreement No. 647554).

# Appendix A. Optical Emission Spectroscopy analysis

The evolution of HiPIMS plasma composition, for $P_{Peak}$ values of 5 kW, 27 kW and 50 kW, is reported in Fig. A1. The $P_{Peak}$ level strongly influences the number of the different species present in the plasma and, thus, the corresponding peak intensities. We carried out a fit, for all the marked peaks in Fig.

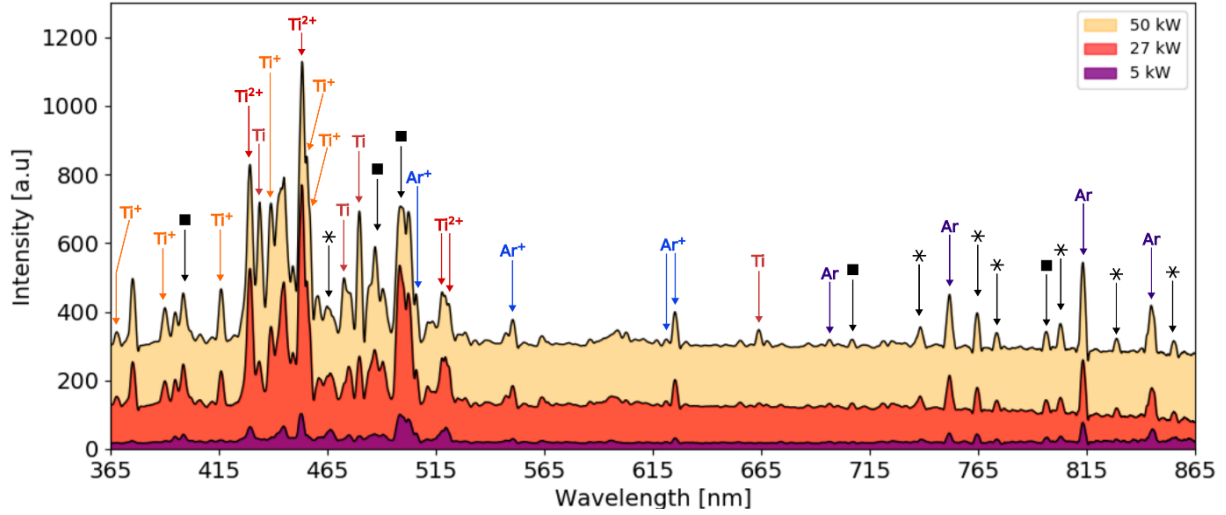

**Fig. A1**. Evolution of the HiPIMS optical emission spectrum for three different values of peak power (5 kW, 27 kW, 50 kW). Arrows indicate the emission lines for Ti, $Ti^+$, $Ti^{2+}$, Ar and $Ar^+$ identified with the procedure described in this appendix and used to evaluate the emission intensity ratios for the HiPIMS regime. Peaks identified by an asterisk belong to Ar or $Ar^+$ species, while peaks identified by a black square belong to Ti, $Ti^+$ or $Ti^{2+}$ species, accordingly only to the NIST database.

A1, to determine the subtended area (i.e. the intensity) for each value of $P_{Peak}$ investigated (we considered a number $n_P$ of $P_{Peak}$ equal to 9, ranging from 5 to 50 kW). The fits were performed with the Levenberg-Marquardt least-square fitting algorithm. As reported in literature [69,70] the experimental peak profiles may be described as the combination of various functions (i.e. Gaussian, Lorentzian and Voigtian profiles) that take into account the internal processes occurring in the discharge. Performing the fit, we considered for the OES peaks a Gaussian shape only, while the local background was modeled with a linear profile.

To associate each peak to the corresponding plasma species, we have considered spectra acquired at $n_P$ different $P_{Peak}$, identified by the $i$-index. In each spectrum we have fitted $j$ peaks associated to the same species to evaluate their intensities $I_{i,j}$.

It is reasonable to assume that peaks associated to the same species will exhibit a characteristic $P_{Peak}$

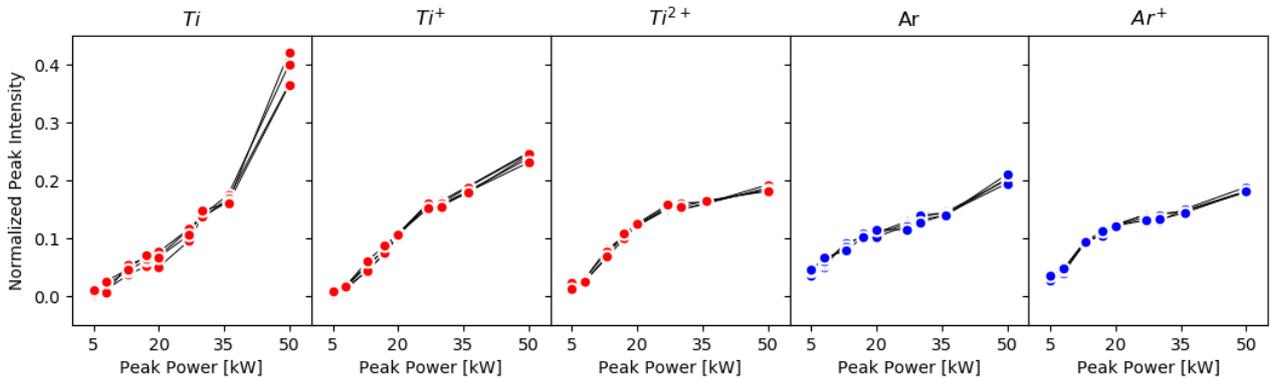

**Fig. A2.** Normalized OES peaks intensity as a function of peak power for Ti, $Ti^+$, $Ti^{2+}$, Ar and $Ar^+$ species.

vs $I_{i,j}$ relationship. If the normalized $I_{i,j}$ over $\sum_{i=1}^{n_P} I_{i,j}$ is considered, the trends associated to one specific species collapse on the same curve as shown in Fig. A2.

This procedure, coupled with data provided by the NIST database [71] allows us to safely assign the considered OES peaks to the different species present during the deposition. We excluded the peaks for which the presented procedure and the NIST data are not in absolute agreement (e.g. caused by the superposition of different lines). Those peaks are indicated with an asterisk and square marker in Fig. A1.

Most of the peaks in the first region of the spectra are referred to neutral Ti and singly charged $Ti^+$ ions, while a few peaks refer to doubly charged $Ti^{2+}$ ions. Moreover, at the end of the region, there are peaks related to singly charged $Ar^+$ ions. As already mentioned, in the second region we detected neutral Ar and $Ar^+$ peaks only. All the identified emission lines and the corresponding excitation energies ($E_k$) are reported in Table A1.

**Table A1.** Identified emission lines and corresponding excitation energies for Ti, $Ti^+$, $Ti^{2+}$, Ar and $Ar^+$.

| Ti | | $Ti^+$ | | $Ti^{2+}$ | | Ar | | $Ar^+$ | |
|---|---|---|---|---|---|---|---|---|---|
| Wavelength [nm] | $E_k$ [eV] | Wavelength [nm] | $E_k$ [eV] | Wavelength [nm] | $E_k$ [eV] | Wavelength [nm] | $E_k$ [eV] | Wavelength [nm] | $E_k$ [eV] |
| 433.8 | 5.017 | 367.98 | 4.95 | 429.25 | 22.62 | 696.5 | 13.32 | 506.1 | 19.26 |
| 472.49 | 3.69 | 390.1 | 4.3 | 453.36 | 20.79 | 751.6 | 13.27 | 550.4 | 25.06 |
| 479.77 | 4.91 | 416 | 4.06 | 517.79 | 18.31 | 813 | 13.07 | 621.18 | 19.26 |
| 663.9 | 5.19 | 438.911 | 4.05 | 521.06 | 18.31 | 844.8 | 13.09 | 625.9 | 19.68 |
| - | - | 455.67 | 3.93 | - | - | - | - | - | - |
| - | - | 457.14 | 4.28 | - | - | - | - | - | - |

# Appendix B. Summary of film depositions and characterizations

Table B1 shows all the samples produced to carry out this research work. For all the samples two deposition parameters (deposition time and substrate bias voltage), thickness and performed characterizations are reported.

Table B1. List of deposited samples, characterizations and thicknesses.

| Deposition Mode | Deposition Time [min] | Bias Voltage [V] | Thickness [nm] | XRD | Stress | Morphology (SEM) | Annealing |
|---|---|---|---|---|---|---|---|
| DCMS | 15.0 | 0 | 78 | ✓ | ✓ | ✓ | ✗ |
| DCMS | 22.5 | 0 | 126 | ✓ | ✓ | ✓ | ✗ |
| DCMS | 30.0 | 0 | 221 | ✓ | ✓ | ✓ | ✗ |
| DCMS | 60.0 | 0 | 446 | ✓ | ✓ | ✓ | ✗ |
| DCMS | 90.0 | 0 | 658 | ✓ | ✓ | ✗ | ✗ |
| HiPIMS | 15.0 | 0 | 51 | ✓ | ✓ | ✗ | ✗ |
| HiPIMS | 30.0 | 0 | 67 | ✓ | ✓ | ✓ | ✗ |
| HiPIMS | 37.5.0 | 0 | 92 | ✓ | ✓ | ✗ | ✗ |
| HiPIMS | 45.0 | 0 | 105 | ✗ | ✓ | ✗ | ✗ |
| HiPIMS | 45.0 | 0 | 128 | ✓ | ✓ | ✓ | ✗ |
| HiPIMS | 60.0 | 0 | 164 | ✓ | ✓ | ✗ | ✗ |
| HiPIMS | 120.0 | 0 | 396 | ✓ | ✓ | ✓ | ✗ |
| HiPIMS | 180.0 | 0 | 580 | ✓ | ✗ | ✓ | ✗ |
| HiPIMS | 15.0 | -300 | 36 | ✓ | ✓ | ✗ | ✗ |
| HiPIMS | 30.0 | -300 | 79 | ✓ | ✓ | ✓ | ✗ |
| HiPIMS | 45.0 | -300 | 110 | ✓ | ✓ | ✗ | ✗ |
| HiPIMS | 60.0 | -300 | 137 | ✓ | ✓ | ✓ | ✗ |
| HiPIMS | 90.0 | -300 | 198 | ✓ | ✓ | ✗ | ✗ |
| HiPIMS | 120.0 | -300 | 258 | ✓ | ✗ | ✓ | ✗ |
| HiPIMS | 180.0 | -300 | 587 | ✓ | ✗ | ✓ | ✗ |
| HiPIMS | 15.0 | -500 | 35 | ✓ | ✓ | ✗ | ✗ |
| HiPIMS | 22.5 | -500 | 66 | ✓ | ✓ | ✗ | ✗ |
| HiPIMS | 30.0 | -500 | 76 | ✓ | ✓ | ✓ | ✗ |
| HiPIMS | 45.0 | -500 | 97 | ✓ | ✓ | ✗ | ✗ |
| HiPIMS | 45.0 | -500 | 141 | ✓ | ✓ | ✓ | ✗ |
| HiPIMS | 60.0 | -500 | 165 | ✓ | ✓ | ✗ | ✗ |
| HiPIMS | 120.0 | -500 | 329 | ✓ | ✓ | ✓ | ✗ |
| HiPIMS | 180.0 | -500 | 495 | ✓ | ✗ | ✓ | ✓ |